\documentclass[conference]{IEEEtran}
\usepackage{cite}
\ifCLASSINFOpdf
  \usepackage[pdftex]{graphicx}
  \DeclareGraphicsExtensions{.pdf,.jpeg,.png}
\else
\fi
\usepackage{amsmath}
\usepackage{algorithmic}
\usepackage{array}
\ifCLASSOPTIONcompsoc
 \usepackage[caption=false,font=normalsize,labelfont=sf,textfont=sf]{subfig}
\else
 \usepackage[caption=false,font=footnotesize]{subfig}
\fi
\usepackage{url}
\usepackage{comment}

\begin{document}
% \title{FPGA Hardware Accelerated Secret Sharing MPC: A COPA Use Case}
% \title{COPA Use Case: Distributed Secure Joint Computation}
\title{Distributed Hardware Accelerated Secure Joint Computation on the COPA Framework}




\author{\IEEEauthorblockN{Rushi Patel\IEEEauthorrefmark{1},
Pouya Haghi\IEEEauthorrefmark{1},
Shweta Jain\IEEEauthorrefmark{3},
Andriy Kot\IEEEauthorrefmark{3},
Venkata Krishnan\IEEEauthorrefmark{3}, \\
Mayank Varia\IEEEauthorrefmark{2} and
Martin Herbordt\IEEEauthorrefmark{1}}
\IEEEauthorblockA{\IEEEauthorrefmark{1}College of Engineering, Boston University, Boston, MA}
\IEEEauthorblockA{\IEEEauthorrefmark{2}Computing \& Data Sciences, Boston University, Boston, MA}
\IEEEauthorblockA{\IEEEauthorrefmark{3}Intel Corporation, Hudson, MA}
Email: 
\IEEEauthorrefmark{1}\{ruship,haghi,herbordt\}@bu.edu 
\IEEEauthorrefmark{2}varia@bu.edu 
\IEEEauthorrefmark{3}\{shweta.jain,andriy.kot,venkata.krishnan\}@intel.com}

\maketitle

\IEEEpubidadjcol

\begin{abstract}
Performance of distributed data center applications can be improved through use of FPGA-based SmartNICs, which provide additional functionality and enable higher bandwidth communication. Until lately, however, the lack of a simple approach for customizing SmartNICs to application requirements has limited the potential benefits. Intel’s Configurable Network Protocol Accelerator (COPA) provides a customizable FPGA framework that integrates both hardware and software development to improve computation and communication performance. 
In this first case study, we demonstrate the capabilities of the COPA framework with an application from cryptography -- secure Multi-Party Computation (MPC) -- that utilizes hardware accelerators connected directly to host memory and the COPA network. We find that using the COPA framework gives significant improvements to both computation and communication as compared to traditional implementations of MPC that use CPUs and NICs. A single MPC accelerator running on COPA enables more than 17Gb/s of communication bandwidth while using only 1\% of Stratix 10 resources. We show that utilizing the COPA framework enables multiple MPC accelerators running in parallel to fully saturate a 100Gbps link enabling higher performance compared to traditional NICs.
\end{abstract}

\IEEEpeerreviewmaketitle

\section{Introduction}

Distributed systems utilize many-node environments and communicate through the use of networks to run large-scale applications with large data sets. Data centers provide ideal environments for distributed computing as they offer low-latency communication between nodes but are often limited by network bandwidth through the use of general-purpose NICs. These network bottlenecks drive the need for alternative communication resources to improve performance of large-scale datacenter applications.
% continue to hold back the potential performance available running in a data center.
SmartNICs \cite{zilberman2014netfpga, mellanoxbluefield, mellanoxinnova2} have been introduced to perform the same  tasks of standard NICs, but contain additional resources to allow for network function optimization with additional hardware. Microsoft has shown \cite{Caulfield16, firestonensdi2018, bojie2016clicknp} the usage of dedicated SmartNICs with FPGA resources for network function offload and cloud management features. Adoption of SmartNICs continues to increase as a means to accelerate network functions and offload packet processing tasks away from CPU resources \cite{LeSoCC2017UNO, Schonbein19, EranATC19NICA, torkASPLOS20Lynx, GrantSIGCOMM20FairNIC, Sebastiano2019DDOS}.

Past avenues of research focus have focused on ASIC-based SmartNICs that utilize soft cores for packet processing, but face challenges in hardware adoption of applications and are limited by network speed. In contrast, SmartNICs with integrated FPGAs address these limitations by offering a reconfigurable environment and enable inline acceleration of network functions. Inline accelerators offer many opportunities to perform packet processing and filtering, however can still be limited in some high performance computing (HPC) applications.
% such as distributed headless accelerators.

SmartNIC accelerated applications generally rely on vendor support in the form of intellectual property (IP) and software development kits. These systems typically are in early stages of development and/or provide only low-level functionality that is inadequate functionality for many use cases. Thus an additional hurdle is the implementation of software application layers which can natively communicate with these SmartNIC resources.

% In contrast Intel's Configurable Network Protocol Accelerator (COPA) \cite{Krishnan20, krishnan2020configurable} provides two options to reconfigurable accelerators, inline and lookaside.
% The framework utilizes the open source software library, OpenFabric interface (OFI) libfabric \cite{OpenFabrics2015}, for platform-agnostic development and a standard for networking and acceleration invocation.
Intel's Configurable Network Protocol Accelerator (COPA) \cite{Krishnan20, krishnan2020configurable} was developed to address these issues. COPA utilizes the open source software library, OpenFabric interface (OFI) libfabric \cite{OpenFabrics2015}, for platform-agnostic development and a standard for networking and acceleration invocation. In addition, the COPA hardware framework provides two options to reconfigurable accelerators, inline and lookaside, both of which are directly accessible from the libfabric API. COPA uses the on-board high speed transceivers, e.g., of the Intel Stratix 10 GX, and a uniquely designed architecture to enable high speed remote direct memory access (RDMA) between nodes at 100Gb/S line rate. Unique features include the ability for remote invocation of accelerators and headless operations for host free integration into a distributed data center environment. So far, however, there has been no published work demonstrating or evaluating COPA with respect to a distributed application; that is our goal here. 

As a candidate application we have selected Multi-Party Computation (MPC), which would greatly benefit from the features available through the COPA framework. MPC is the cryptographic process of performing calculations on confidential data between multiple organizations while maintaining a level of confidentiality, integrity, and assurance of one's own private data. Parties encode and share their own private data between organizations while maintaining an agreed upon level of security guarantee. This form of joint computation is especially important for industries such as healthcare and finance, as user data is typically under protection through laws and regulations. FPGA accelerated Multi-Party Computation continues to be a progressive research topic \cite{Jarvinen2010, Jarvinen2010b, Frederiksen2014, songhori16, hussain18, songhori19, hussain19, Fang2017a, Fang2019, Huang2019a, Leeser2019, Wolfe20, Patel20} as significant performance improvements can be obtained from hardware acceleration.

This paper argues that combining the COPA tool-set with state of the art MPC algorithms can achieve a lower communication bottleneck for high performance computation inside a datacenter environment. We show that utilizing the COPA system enables a method of performing low-level MPC operations with minimal CPU interaction while enabling improved performance compared to traditional CPU and NIC implementations.

In summary, our contributions are as follows:
\begin{itemize}
    \item Examine the performance available utilizing the 100Gbps network and configurable lookaside accelerator option of the COPA FPGA framework.
    \item Adapt hardware accelerated MPC operations to the COPA infrastructure utilizing the configurable FPGA lookaside accelerator enabling significant performance improvements compared against CPU and NICs.
    \item Using only 1\% of the FPGA fabric for secure joint computation, we show that having multiple accelerators running in parallel can saturate the potential 100Gb/s link available through COPA while performing over 6000 MPC operations per second.
\end{itemize}

%%%%%%%%%%%%%%%%%%%%%%%%%%%%%%%%%%%%%%%%%%%%%%%%%%%%%%%%%%%%%%%%%%%%%%%%%%%%%%%
\section{Background}

\subsection{Configurable Network Protocol Accelerator}

The COPA SmartNIC works by using the libfabric software layer with included extensions to queue commands for processing by the hardware. These commands include both RDMA functionality and accelerator specific commands. The RDMA functions use the COPA network TX and RX data paths to perform memory read and write functions without host involvement. Previous work has shown the COPA network can achieve up to 100Gb/S bandwidth with zero-copy direct memory access.

The accelerator specific commands include the use of a number of inline accelerators alongside the RDMA functions. Inline functions operate on packets during transit in a bump-in-the-wire method, allowing for data manipulation of packets during egress or ingress of edge nodes. Two examples of inline acceleration are checksum calculation on data in transit or encode/decode functions with pre-shared key pairs.

Additionally, commands can be constructed to trigger the lookaside accelerator functions both locally and on remote COPA nodes. Lookaside functions operate on data at rest, in host memory, and have direct connection with the COPA network to perform additional RDMA functions. This enables the lookaside accelerator to perform data transfer tasks without requiring the host to initiate network operations. Both inline and lookaside accelerator options can be reconfigured by users for application specific functions. 
%We envision the usage of these accelerators in two methods, either developed in-house by application designers or available as an FPGA-as-a-Service system where accelerators are programmed before tenants request usage.

\subsection{Secure Multi-Party Computation}

Secret sharing-based multi party computation (MPC) is a method of secure joint computation that allows any number of party members $N$ to work together to obtain a final output \cite{araki16,DBLP:conf/ndss/Demmler0Z15,dalskov2021fantastic}.  Confidential data is distributed to all party members in the form of \emph{shares}, where each individual share does not contain enough information to learn anything about the secret data but all shares collectively can be used to decode the data. Each party member is considered an equal to another, and computation is performed synchronously between all members to maintain accurate share representations of the final value between all members. Each synchronous operation requires members to perform communication rounds of data among all parties, which increases the need for high bandwidth and low latency networks between members. As a consequence, the rate of computation in MPC is bottlenecked by network performance, making this application a prime target for COPA.
% Once computation is finalized, each party member will hold a share of the final output, and party members reveal the final value through one last round of communication between members.
% If computation is performed correctly, the final output will be accurate based on initial inputs while never revealing the data provided by the party members.

Secret sharing MPC requires all party members to perform both local and joint computation. With FPGAs secret sharing MPC can obtain performance improvements on both forms of computation tasks \cite{Wolfe20, Patel20}. Prior research shows that a single FPGA can fully saturate a standard 10GigE NIC while only utilizing a fraction of the available resources of the FPGA hardware. In addition, colocated party members in a datacenter provides an optimal environment to reduce communication latency further improving the performance of communication-dependent MPC operations.

To the best of our knowledge, we are the first to integrate Multi-Party Computation and SmartNIC functionality to improve upon communication bottlenecks. We believe the combination of MPC and COPA lookaside acceleration enables a significant improvement to both computation and communication performance thus eliminating previous network limitations. Utilizing the many unique features of the COPA framework, including remote accelerator triggering and payload processing before and after transit, allow for a headless behavior of many MPC operations between party members. This enables less CPU utilization during concurrent operations between party members and reduces the need for explicit synchronized commands by each host system.

%%%%%%%%%%%%%%%%%%%%%%%%%%%%%%%%%%%%%%%%%%%%%%%%%%%%%%%%%%%%%%%%%%%%%%%%%%%%%%%

\begin{figure}
    \centering
    %\resizebox{\linewidth}{!}{\input{tikz/and_1}}
    \includegraphics[width=0.9\linewidth]{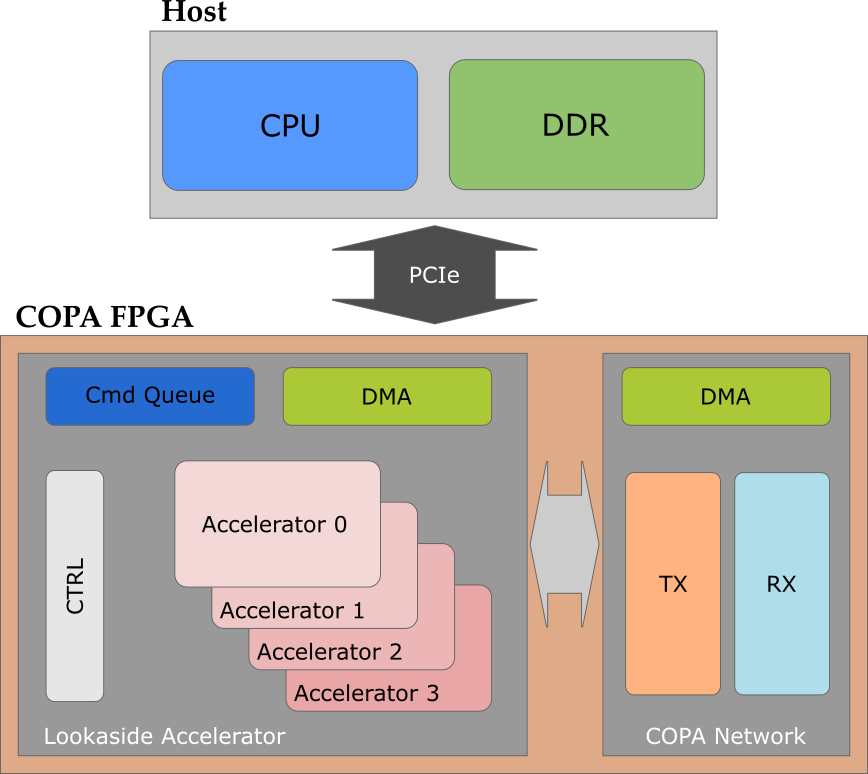}
    \vspace*{-0.1truein}
    \caption{COPA architecture connected to host system through PCIe. The COPA FPGA contains a lookaside accelerator implementation directly connected to the COPA network allowing for network functions without host interaction.}
    \label{fig:MPC_LA}
    \vspace*{-0.1truein}
\end{figure}

\begin{figure}
    \centering
    %\resizebox{\linewidth}{!}{\input{tikz/and_1}}
    \includegraphics[width=0.6\linewidth]{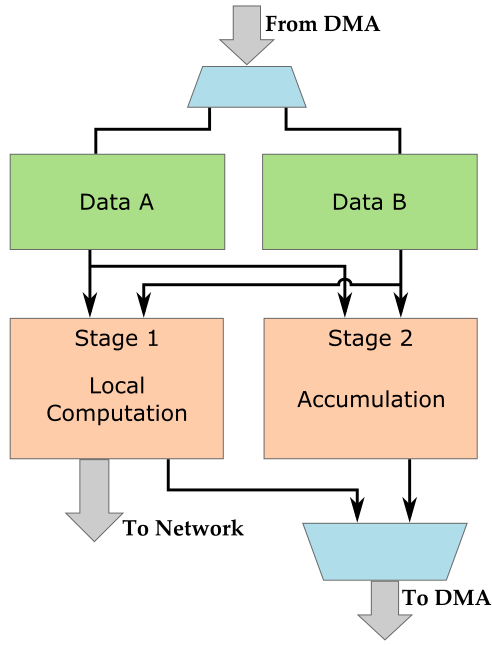}
    \vspace*{-0.1truein}
    \caption{Two-stage MPC implementation as single lookaside accelerator.}
    \label{fig:MPC_LA_Accel}
    \vspace*{-0.2truein}
\end{figure}

%%%%%%%%%%%%%%%%%%%%%%%%%%%%%%%%%%%%%%%%%%%%%%%%%%%%%%%%%%%%%%%%%%%%%%%%%%%%%%%
\section{Architecture Implementation}

\subsection{Initial Protocol Requirements}

Our initial design uses the host to generate, and COPA unidirectional PUT functions to distribute, shares of private data to the other parties. Each share is generated through of the use of a random number generator and a specific calculation following the agreed upon algorithm protocol. In our implementation, we use the 4-party MPC protocol of Dalskov, et al. \cite{dalskov2021fantastic}; we discuss algorithm specifics in Section \ref{Experiment_Setup}. As all confidential data remains tied to the host, this maintains full privacy of confidential data as any data passing through the COPA network is obfuscated in the form of shares.
% Each party does not contain enough information to determine the clear-text value but requires information from other party member to reconstruct the share.

In addition to distributing shares between parties, a set of keys are also allocated for further computation when pseudorandomly generated numbers must be known by two or more parties.
% Our pseudo random number generator (PRNG) utilizes the OpenCores projects \cite{Homer2012, Castillo2004} IP for the design.
Each party member uses a set of unique keys to generate random numbers concurrently with other party members; however, each party member does not contain knowledge of all keys. This concurrent behavior is important to maintain protocol accuracy between party members during operation and can help avoid additional communication.

\subsection{MPC Gate Operations}

We focus on arithmetic MPC operations which consist of addition and multiplication.
Shares are formed using \emph{128-bit} values and all operations are based around modular arithmetic with a ring module of ${2^{128}}$.
% Functionality for Boolean shares is similar to arithmetic and our implementation allows for this additional operation with a simple selector input to change forms.
For our algorithm of choice, MPC addition consists of local computation only and doesn't require interaction between members.
% This means shares of values can be directly computed on by each party member and the output remains accurate for future computation.
Multiplication requires both local computation of shares and a single round of communication between party members; we thus focus on the performance obtained by improving on these communications.

\subsection{Lookaside Accelerator}

The COPA lookaside architecture uses separate accelerator logic outside of the COPA network as seen in Figure \ref{fig:MPC_LA}. Acceleration is initially controlled by the host through a unique command containing the source data location, destination location, length of data, and type of operation. A global control unit manages incoming commands in the queue and assigns them to appropriate accelerators. This feature enables a single host to issue commands to different accelerators for added parallelism or unique functions. We use only a single MPC accelerator for our initial tests, but discuss the improvements available with additional accelerators. Each accelerator initially collects the source data through a DMA operation from the host memory. If source data is unavailable locally, i.e., found on a remote COPA node, then a COPA network command is generated and used to obtain data from the remote node prior to DMA operation. Following the local DMA completion, acceleration is performed and final calculations are sent back to host memory through a second DMA operation. If the destination memory location is for a remote node, then the COPA network is again used to send the final completed values to another COPA node on the network.

\subsubsection{MPC Accelerator Core}

To fully utilize the COPA lookaside accelerator we split up the multiplication operations into the local computation stage and a post communication accumulation stage as seen in Figure \ref{fig:MPC_LA_Accel}. Data is first obtained by the DMA logic and stored into two sets of on-chip memory, Data A and Data B. Data in these two on-chip memory regions are based on the stage of computation being performed. First stage computation takes two lists of values in share form, while the second stage contains the intermediate share data and communication data received from other party members. The local computation stage uses a pseudo random number generator and on-chip resources to perform the initial MPC calculations and save the intermediate share information back to host memory. Addition operations only require the use of the first stage accelerator to perform calculation on input data and generate complete shares. Multiplication operations use the first stage to prepare partial local shares and data for communication to other party members. 

On completion of the first stage, information required for party members is prepared for communication and passed along appropriately through the use of the COPA network.
This prepares the data for processing in the second stage of the multiply operation.
On completion of communication from each party member, the locally generated intermediate shares are combined with ingress data from party members and saved back into host memory for future computation. 

%%%%%%%%%%%%%%%%%%%%%%%%%%%%%%%%%%%%%%%%%%%%%%%%%%%%%%%%%%%%%%%%%%%%%%%%%%%%%%%
\section{Results}

\begin{figure}
    \centering
    \includegraphics[width=0.9\linewidth]{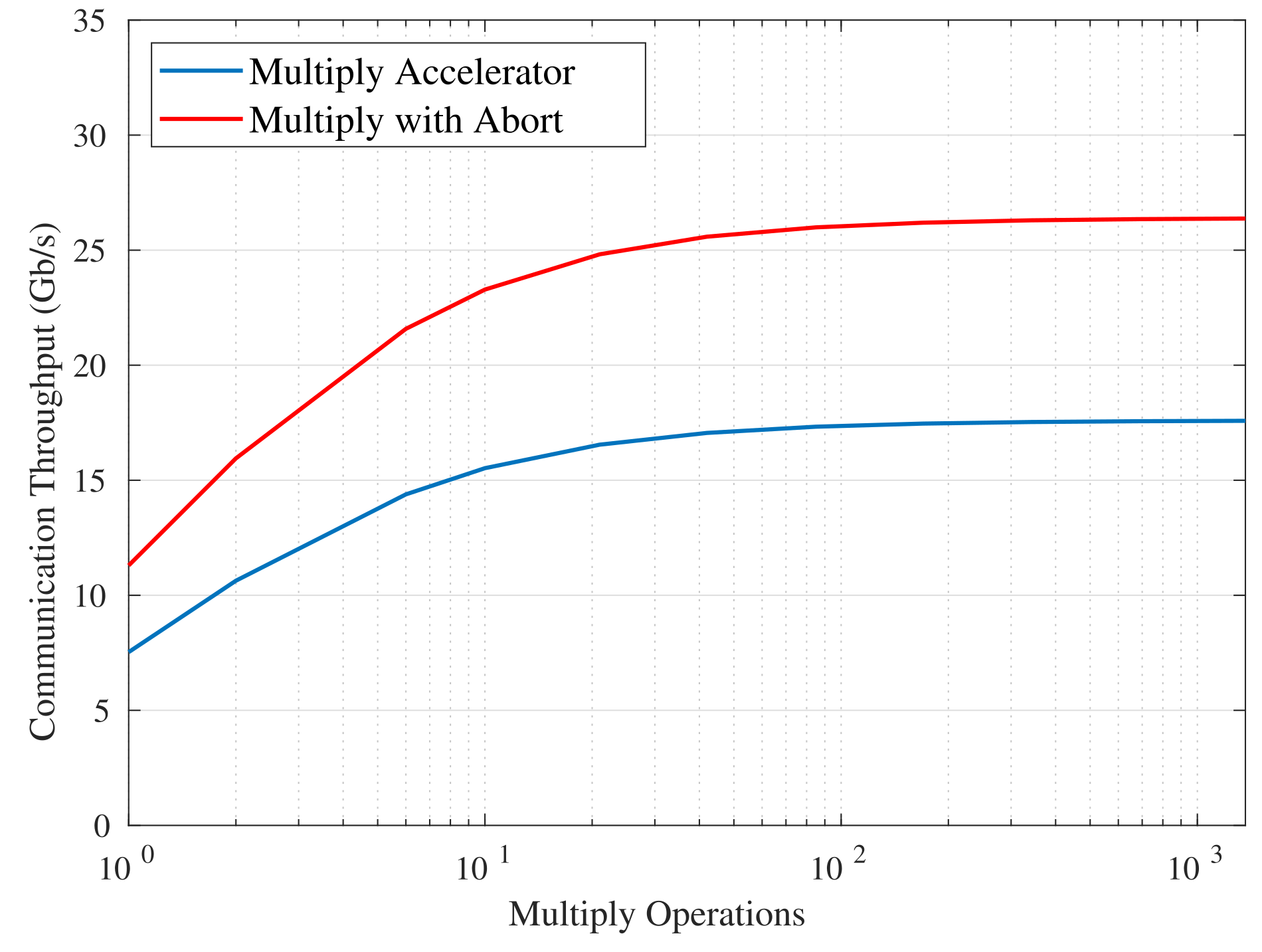}
    \vspace*{-0.1truein}
    \caption{Throughput comparison between base MPC accelerator and MPC with malicious security running at 275 MHz with varying batch sizes of multiply operations. Malicious security requires an additional collision resistant hash value to be transferred. Saturation of the base accelerator is over 17.5Gb/s and with malicious security is over 26.3Gb/s.}
    \label{fig:MPC_Throughput}
    \vspace*{-0.1truein}
\end{figure}

\begin{table}
\centering
\caption{Single MPC lookaside FPGA utilization}
\vspace*{-0.05truein}
\begin{tabular}{|c|c|}
\hline
Stratix 10 FPGA & Raw (Total Percentage) \\
\hline\hline
Freq & 250MHz - 275MHz \\
\hline
ALM & 10667 (1\%)\\
\hline
Memory bits & 5,156,500 (2\%)  \\
\hline 
RAM blocks & 668 (6\%)  \\
\hline 
DSP blocks & 150 (3\%)  \\
\hline 
\end{tabular}
\label{fig:FPGA_Utilization}
\vspace*{-0.2truein}
\end{table}

\subsection{Experiment Setup}
\label{Experiment_Setup}

We focus on the throughput and resource utilization of the multiply operation for a 4-party semi-honest majority MPC algorithm \cite{dalskov2021fantastic}. Our chosen algorithm requires each party member to hold 3-out-of-4 shares of each data element and communicate between all parties equally during calculation. In particular, the key is performing multiplication operations on shares of \emph{128-bit} integer data types which generates three \emph{128-bit} integers for communication to the other party members. 

To maintain accuracy in multiply operations, each party member must perform the initial local calculations synchronously and participate in a round of communication to obtain all the necessary intermediate data from other party members. For additional security against malicious parties, hashed values of communication data is sent to opposing parties as a method of verification between party members. A mismatch in calculated hashes would indicate an incorrect joint computation and the party members can abort future computation to avoid the risk of data leakage through a malicious actor.

We implement our hardware design on the COPA framework using Intel Stratix 10 FPGAs interconnected with 100GigE high speed transceivers. Each party maintains ownership of a single FPGA connected to a host system using the COPA framework for communication between party members. Acceleration is performed through the use of lookaside accelerator commands sent from each host system directly to the FPGA lookaside accelerator through a dedicated queue. The lookaside command format allows for batch operations on a stream of data from a specified source and saves local computation back to host memory while preparing the network data for transfer to each party member.

\subsection{Analysis}

Resource utilization for a single MPC lookaside accelerator can be found in Table \ref{fig:FPGA_Utilization}. This shows the implementation uses minimal resources which allows for the inclusion of more accelerators into each COPA FPGA; these additional accelerators may include multiple instances of the MPC core operations or additional functionality for High Performance Computing applications such as collectives.
%(e.g. Scatter Gather, Map Reduce \cite{Haghi20a}).
% At this time, we are limited to a maximum frequency of $275MHz$.
With the inclusion of a single MPC accelerator, Figure \ref{fig:MPC_Throughput} shows how much data is available for communication based on the input length of the lookaside accelerator source data.

The pipeline implementation of the accelerator allows for data to be processed and ready for communication every cycle, after an initial startup delay accessing host memory. Using a single accelerator and batching multiplication operations over a stream of source data, the accelerator performs enough computation to saturate a traditional 10Gb/s link. These results are similar to past implementations \cite{Wolfe20, Patel20} and show that integration with the COPA system is beneficial to improve the total throughput possible with these hardware implemented MPC operations.

Examining the throughput of large batches of multiplication operations, Figure \ref{fig:MPC_Throughput} shows a single accelerator performing the basic algorithm (without abort) can saturate a 17.5Gb/s connection, while the inclusion of additional malicious security for abort requires larger than 26.3Gb/s connection to avoid saturation. We can therefore include up to 6 MPC accelerators without abort, or 4 MPC accelerators with abort, to saturate the COPA network. 

In addition to the communication improvements, the COPA system enables a set-and-forget method for acceleration and communication which frees up each host processor to perform additional non-MPC functions. Queuing operations into the lookaside accelerator, with knowledge that data will be shared appropriately, allows for final completion of each operation without the need to block the process on each step.

%%%%%%%%%%%%%%%%%%%%%%%%%%%%%%%%%%%%%%%%%%%%%%%%%%%%%%%%%%%%%%%%%%%%%%%%%%%%%%%
\section{Conclusion}

The COPA framework enables hardware acceleration and improved network functions for, potentially, many different applications. We show how an MPC implementation fits into the COPA framework and enables improvements to both computation and communication by using the lookaside accelerator features and improved network data transfer. In addition, MPC running on the COPA system enables the use of the an open-source software library, OFI, as an alternative to specialized MPC software used by each party member. We aim to increase the size of our tests with additional MPC algorithms aiming for two party, three party, and four party computation alternatives. We also aim to enable more accelerator options to improve on secure joint computation through the means of memory operations such as scatter/gather. This will enable additional improvements to both MPC throughput and network communication.

%%%\subsection{Future Work}

Our future work will include a full end-to-end method of MPC using the COPA hardware/software infrastructure for cloud/data centers, MPC-as-a-Service. 
% This work will utilize many security features available in modern cloud environments to create dedicated MPC networks preventing attackers from snooping on MPC operations.
% We also aim to make the usage of MPC in the cloud as easy as possible to remove the large barrier to entry currently involved with running MPC with multiple parties.
Fully utilizing the COPA features, we aim to create a fully autonomous MPC system allowing for any host to perform trusted operations on the collective data. With COPA remote invocation of target accelerator nodes, we aim to enable complete MPC applications to run with only a single host triggering the operation and each additional party member acting like a headless target node. By using all of these features MPC-as-a-Service can be a viable method of trusted secure joint computation with a minimal barrier to entry.

% \section*{Acknowledgment}



\bibliographystyle{IEEEtran}
\bibliography{main.bib}

\end{document}